\documentclass[english]{article}
\pdfoutput=1
\usepackage[T1]{fontenc}
\usepackage[latin9]{inputenc}
\usepackage{units}
\usepackage{textcomp}
\usepackage{amsmath}
\usepackage{amssymb}
\usepackage{graphicx}
\usepackage{esint}

\makeatletter
\newcommand{\lyxaddress}[1]{
\par {\raggedright #1
\vspace{1.4em}
\noindent\par}
}

\usepackage{babel}

\makeatother

\usepackage{babel}
\begin{document}

\title{Lagrangian Numerical Methods for Ocean Biogeochemical Simulations}

\author{Francesco Paparella$^{1,2}$\\
Marina Popolizio$^{2}$}

\maketitle

\lyxaddress{$^{1}$Division of Sciences and Mathematics\\
$\hphantom{^{1}}$New York University Abu Dhabi}

\lyxaddress{$^{2}$Dipartimento di Matematica e Fisica\\
$\hphantom{^{2}}$Università del Salento Lecce}
\begin{abstract}
We propose two closely--related Lagrangian numerical methods for the
simulation of physical processes involving advection, reaction and
diffusion. The methods are intended to be used in settings where the
flow is nearly incompressible and the Péclet numbers are so high that
resolving all the scales of motion is unfeasible. This is commonplace
in ocean flows. Our methods consist in augmenting the method of characteristics,
which is suitable for advection--reaction problems, with couplings
among nearby particles, producing fluxes that mimic diffusion, or
unresolved small-scale transport. The methods conserve mass, obey
the maximum principle, and allow to tune the strength of the diffusive
terms down to zero, while avoiding unwanted numerical dissipation
effects.
\end{abstract}

\section{Introduction}

Biogeochemical problems in oceanography are usually expressed in terms
of coupled advection-reaction-diffusion equations involving scalar
fields, sometimes in large number, representing chemical species,
biological species, or functional groups (see, e.g., \cite{Vichi07}).
These fields are advected by the ocean currents, are subject to diffusion,
and interact nonlinearly with each other.

A generic, abstract form of oceanographical biogeochemical equations
is the following 
\begin{align}
\frac{\partial c_{1}}{\partial t}+\boldsymbol{u}\cdot\nabla c_{1}= & D_{1}\nabla^{2}c_{1}+f_{1}(c_{1},\ldots,c_{n})\nonumber \\
\vdots\label{eq:the_problem}\\
\frac{\partial c_{n}}{\partial t}+\boldsymbol{u}\cdot\nabla c_{n}= & D_{n}\nabla^{2}c_{n}+f_{n}(c_{1},\ldots,c_{n})\nonumber 
\end{align}
where $c_{1},\ldots,c_{n}$ are the scalar fields, $\boldsymbol{u}$
is the water velocity field in the region of interest, which is assumed
to be known, $D_{1},\ldots,D_{n}$ are the diffusion coefficients,
and the functions $f_{1},\ldots,f_{n}$ specify the local interactions
among the scalar fields.

The relative importance of the transport and diffusion terms is quantified
by the Péclet numbers 
\[
Pe_{l}=\frac{UL}{D_{l}}
\]
where $U$ and $L$ are, respectively, a characteristic speed and
a characteristic length associated to the velocity field $\boldsymbol{u}$.
The relative importance of the transport and reaction terms is quantified
by the Damköhler numbers 
\[
Da_{l}=\frac{L}{U\tau_{l}}
\]
where $\tau_{l}$ is a characteristic time scale associated with the
reaction described by $f_{l}$.

The Damköhler number for phytoplankton may range from negligibly small
up to $O(10)$ \cite{Pasquero05}. While large values of the Damköhler
number may amplify the patchiness of a reacting scalar as compared
to a non--reacting one \cite{Mahadevan02,Richards2006} and make the
problem stiff, the true source of numerical difficulties in biogeochemical
applications lies in the enormous size of the Péclet number.

If one takes the diffusivities to be the molecular ones (or computed
from the mean square displacement of trajectories of individual plankton
cells) then the Péclet numbers may easily exceed $10^{10}$. Such
a large value is reflected in the fact that ocean tracers (temperature,
salinity, etc.) show structures from the scale of ocean basins down
to submillimetric scales. Even accounting for a continuing rapid pace
of improvement in computer technologies, it is quite obvious that,
in the foreseeable future, no numerical code will be able to resolve
such a wide interval of scales.

In the absence of reaction terms, a reasonable way to deal with unresolved
small scales is to parameterize the advective fluxes due to the unresolved
scales with diffusion operators (often in a more complicated form
than simple Laplacians). To this end there is an impressive array
of techniques, ranging from explicitly adding new terms to the equations
(e.g. in turbulence closures), to using flux or slope limiters (e.g.
in finite volume methods), to advection and interpolation (e.g. in
semi-lagrangian methods) or dealiasing and filters (e.g. in pseudo--spectral
methods). A review of numerical methods used for geophysical flows
is given in \cite{Durran2010}. In all these cases, however, the strength
of the diffusive terms is determined not just by the physical parameters
of the problem, but also by the size of the mesh. In fact, all these
techniques may be viewed as different ways to average out the subgrid
scales. Thus, in the presence of unresolved small scales, the values
of the scalar fields at each grid node must be understood not as a
pointwise evaluation of a function, but as an average over a spatial
region having an extension comparable with the size of a computational
mesh.

Early studies already showed that changing the strength of the diffusive
fluxes representing the unresolved scales may have a dramatic impact
on the reaction terms \cite{Pasquero05,Brentnall03} and warned that
a ``mean field'' approach might be inappropriate for modeling plankton
dynamics. Later studies, conducted using realistic ocean models, showed
strong fluctuations in plankton productivity depending on the advection
scheme used and, most importantly, on the resolution \cite{Levy01,Levy12,Levy13}.
The most recent assessment of the importance of the unresolved structures
is found in \cite{Martin15}.

As a first step to understand these results we need to observe that,
for the full set of equations (\ref{eq:the_problem}), one faces the
overwhelming difficulty that an averaging operator $\overline{}$
does not commute with nonlinear reaction terms: $f_{l}(\overline{c_{1}},\ldots,\overline{c_{n}})\neq\overline{f_{l}(c_{1},\ldots,c_{n})}$.
Because reactions are formally pointwise one would need to compute
$\overline{f_{l}(c_{1},\ldots,c_{n})}$, but all that current grid--based
codes can do is to compute $f_{l}(\overline{c_{1}},\ldots,\overline{c_{n}})$.
The wide chasm of unresolved scales means that the mesh-averaged values
$\overline{c_{1}},\ldots,\overline{c_{n}}$ may be substantially different
from their pointwise counterpart $c_{1},\ldots,c_{n}$. As we shall
see in the following, the bias produced by this effect may have either
sign, depending, among other things, on the initial conditions.

In the absence of any diffusive effect, that is, setting $D_{1,\ldots,n}=0$
in (\ref{eq:the_problem}), it is arguably better to avoid any discretization
involving an Eulerian grid, and use a straightforward implementation
of the method of characteristics. This leads to the following simple
Lagrangian numerical scheme: we uniformly seed the domain $\Omega$
with $M$ particles, having position $\boldsymbol{x}_{i}$, $i=1,\ldots,M$,
and then numerically solve 
\begin{equation}
\begin{cases}
\dot{\boldsymbol{x}}_{i} & =\boldsymbol{u}(\boldsymbol{x}_{i},t)\\
\dot{c}_{1;i} & =f_{1}(c_{1;i},\ldots,c_{n;i})\\
 & \vdots\\
\dot{c}_{n;i} & =f_{n}(c_{1;i},\ldots,c_{n;i})
\end{cases}\label{eq:ODEs}
\end{equation}
with one among many viable ODE solvers. Here and in the following
we use the shorthand notation $c_{l;i}=c_{l}(\boldsymbol{x}_{i},t)$
for the scalars sampled at the location of each particle. It is important
to appreciate that, even when the number of particles is too small
to fully sample the small-scale structures present in the full solution
of the PDEs, the values $c_{l;i}$ remain unaffected by the sparsity
of the sampling, and are only affected by inaccuracies in the solution
of the ODEs (\ref{eq:ODEs}), due, e.g., to an imperfect knowledge
of the velocity field $\boldsymbol{u}$. This scheme is thus immune
from the averaging problem discussed above. If, as is the case in
oceanographic applications, the velocity field $\boldsymbol{u}$ is
divergenceless, or nearly so, then an initially uniform sampling will
remain uniform, or nearly so, at all future times. In this context
the lack of a structured grid is just a nuisance: diagnostic and data
analysis tasks may be performed after resampling the numerical solutions
of (\ref{eq:ODEs}) on a regular grid of choice, using, e.g., the
methods discussed in \cite[§5.3]{Hockney88}.

Unfortunately, the method of characteristics is not directly applicable
to biogeochemical problems: the complete absence of diffusive effects
in (\ref{eq:ODEs}) would lead to paradoxical effects. For instance,
if a water mass containing some phytoplankton but poor of nutrients
were brought close to water masses devoid of phytoplankton but nutrient--rich,
fluxes associated to small--scale motions would seed some plankton
in the nutrient--rich water masses, leading, if the conditions are
right, to a bloom. With the scheme (\ref{eq:ODEs}) a particle full
of phytoplankton could be brought arbitrarily close to a particle
full of nutrients and yet there would be no exchanges between the
two: the plankton would wither, and the nutrients would remain unused.

In this paper we show how to augment the simple Lagrangian scheme
(\ref{eq:ODEs}) with couplings among nearby particles designed to
mimic diffusive effects or, more generally, fluxes due to small-scale,
unresolved transport processes. In order to be acceptable, such a
coupling must possess the following three properties 
\begin{enumerate}
\item respect mass conservation; 
\item obey the maximum principle; 
\item allow to recover the scheme (\ref{eq:ODEs}) in the limit $D_{l}\to0$. 
\end{enumerate}
The importance of mass conservation is fairly obvious. Even for models
using non--conserving reaction terms, there is no reason to introduce
uncontrollable numerical sources and sinks of scalars. Schemes that
do not obey the maximum principle may create maxima and minima unbounded
by the maxima and minima of the initial conditions. In particular,
scalar fields that should be non-negative (e.g. the concentration
of a chemical species) may locally develop negative values, which,
in turn, yield meaningless results with most reaction models. Being
able to recover the scheme (\ref{eq:ODEs}) means that one is free
to tune the strength of the diffusive effects on the basis of modeling
considerations alone, and not because of numerical requirements. We
propose two distinct couplers that satisfy all these three properties.
Of the two methods that we propose, the first is based on an integral
formulation, the second is an heuristic recipe based on physical considerations.
The two methods are distinct in the way used to enforce mass conservation.
In both cases, however, the maximum principle is a direct consequence
of the fact that the concentration of each particle after a diffusive
step is determined as an average involving the concentrations of nearby
particles. Free parameters, appearing in both methods, can be use
to tune the strength of the diffusive effects to extremely low values,
or to zero, thereby maintaining the particles uncoupled.

Particle--based methods are not a novelty. Smoothed particle hydrodynamics
(SPH) has proved to be very suitable for highly compressible astrophysical
problems, but flexible enough to be applied in many other settings
\cite{Monaghan92}, including heat conduction \cite{Monaghan99}.
However, we felt that achieving all three of the above properties
might be not straightforward with an SPH--inspired approach, therefore
our methods are not based upon the differentiation of a smooth kernel.
Other particle--based methods, closer to the spirit of the present
work, have been proposed for diffusion and advection--diffusion equations
\cite{Degond89,Degond90}, but did not gain a large popularity.

Few are the instances in which Lagrangian methods have been applied
to geophysical problems. Nearly all numerical ocean models use grid--based
methods, with the notable exception of the so--called ``slippery
sack'' model \cite{Haertel02}. This was initially a purely adiabatic,
Lagrangian scheme, which was later augmented with a diffusive coupling
between nearby particles \cite{Haertel09}. The Lagrangian scheme
(\ref{eq:ODEs}) has been successfully applied to explain some incongruences
between ecological models and observations \cite{Koszalka07}. When
augmented with a diffusive coupling it has been used to explain the
Fourier spectrum of a plankton concentration field \cite{Bracco09}.
We are not aware of other applications of Lagrangian schemes to ocean
biogeochemistry. There exists more work on Lagrangian methods for
modeling the atmosphere. In particular, a method based on contour
advection and surgery has been highly successful in reproducing the
observed distribution of stratospheric ozone \cite{Edouard96,Mariotti00}.
For a recent survey on Lagrangian methods in atmospheric sciences
see \cite{Lin13}.

\begin{figure}
\begin{centering}
\includegraphics[width=0.5\textwidth]{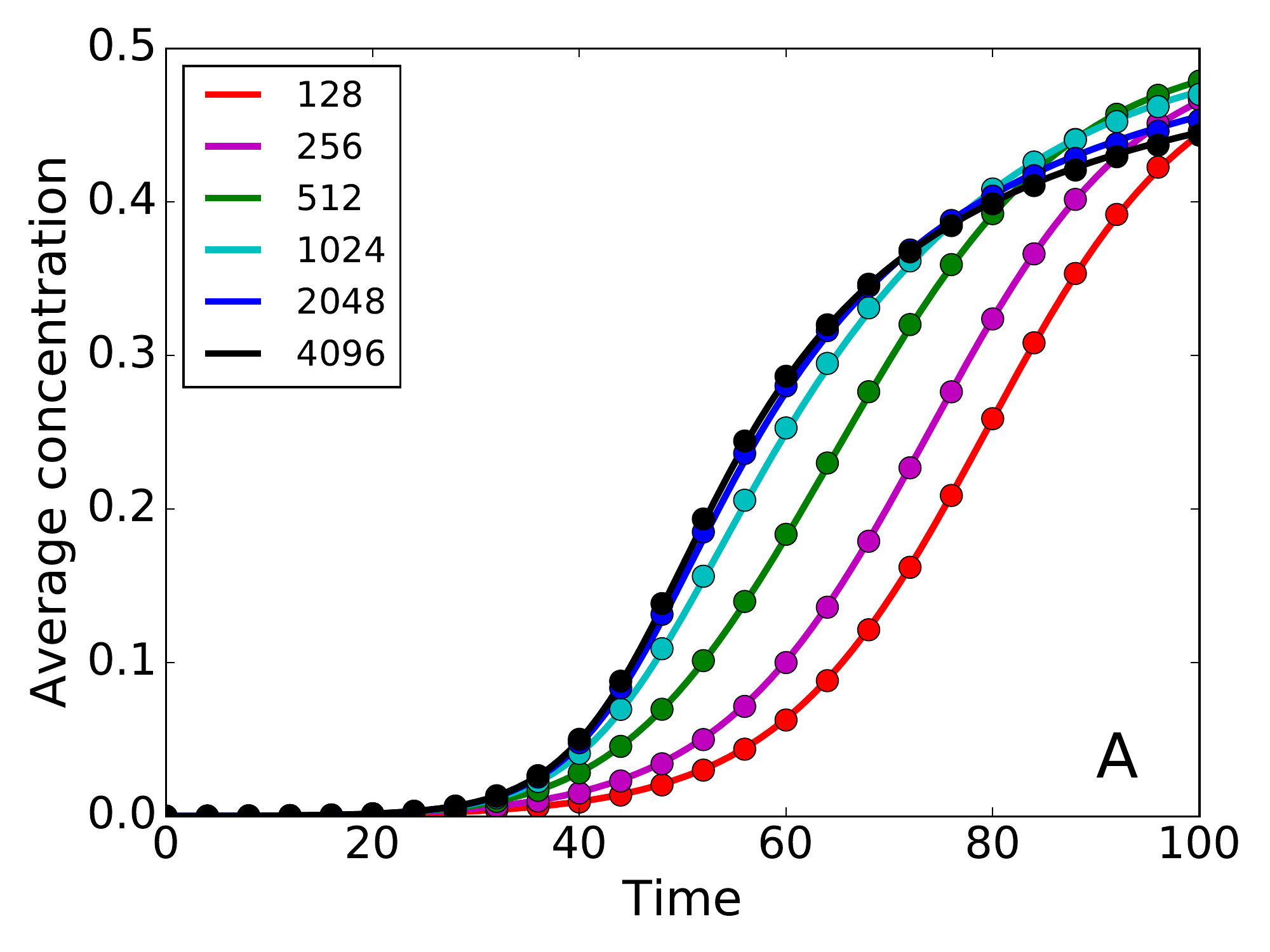}\includegraphics[width=0.5\textwidth]{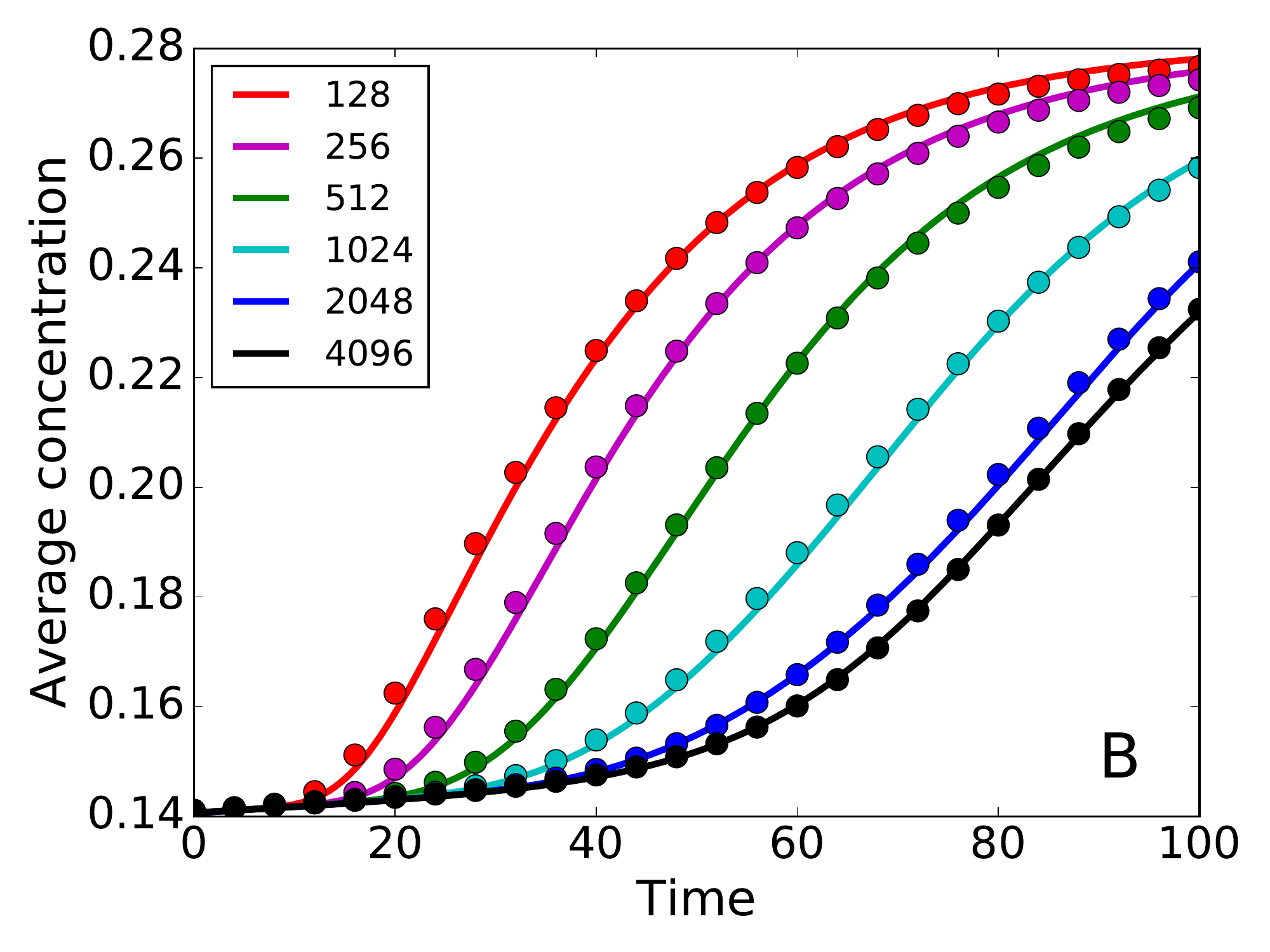} 
\par\end{centering}

\caption{\label{fig:ConsumerProductivity}Spatial average of the field $c_{2}$
as a function of time. The solid lines are results obtained with a
pseudo--spectral code, with progressively higher resolution and correspondingly
lower diffusivity (see text). The dots are results obtained with a
Lagrangian code using the coupler of sec. \ref{sub:Second-coupler}
with $128^{2}$ particles, where the strength of the diffusive coupling
between particles is set as to match that of the pseudo--spectral
computations. Panel A: calculations starting from the initial condition
(\ref{eq:initial_conditions_A}). Panel B: calculations starting from
the initial condition (\ref{eq:initial_conditions_B}).}
\end{figure}

The enormous potential of diffusively--coupled Lagrangian methods
in biogeochemistry is illustrated by a simple example, inspired by
the results obtained with a much more realistic model in \cite{Martin02}.
In (\ref{eq:the_problem}) we set $n=2$ and choose a two--dimensional,
incompressible velocity field $\boldsymbol{u}=(-\psi_{y},\psi_{x})$
defined through the streamfunction $\psi(x,y)=\sin(x)\sin(y)$ on
the doubly--periodic domain $(x,y)\in[0,2\pi)\times[0,2\pi$). The
reaction terms are 
\begin{equation}
f_{1}(c_{1},c_{2})=-r\,c_{1}c_{2},\quad f_{2}(c_{1},c_{2})=+r\,c_{1}c_{2},\label{eq:simple_chemistry}
\end{equation}
with $r=0.2$. We may see the scalar field $c_{2}$ as the spatial
density of a consumer that grows at the expense of a resource whose
density is $c_{1}$. The initial conditions are: 
\begin{equation}
c_{1}(x,y,0)=\cos^{2}(x/2),\quad c_{2}(x,y,0)=10^{-4}.\label{eq:initial_conditions_A}
\end{equation}
We compute six solutions of this problem for progressively smaller
diffusivities and correspondingly higher resolutions. The six meshes
have $128\cdot2^{k}$ points in each direction, and the diffusivities
are $D_{1}=D_{2}=10^{-3}\cdot2^{-2k}$, $k=0,\ldots,5$. At each resolution,
using substantially lower diffusivities would lead to severe oscillations
and numerical instabilities. The solid lines in Figure \ref{fig:ConsumerProductivity}A
show the time evolution of the spatial average of $c_{2}$ (that is,
the mean consumer density). The dots show the same quantity computed
by using the Lagrangian scheme (\ref{eq:ODEs}) augmented with one
of the two diffusive couplers that will be presented in the following
(namely, that of section \ref{sub:Second-coupler}). The six Lagrangian
solutions all use just $128^{2}$ particles, and they differ only
in the strength of the diffusive coupling.

\begin{figure}
\begin{centering}
\includegraphics[width=1\textwidth]{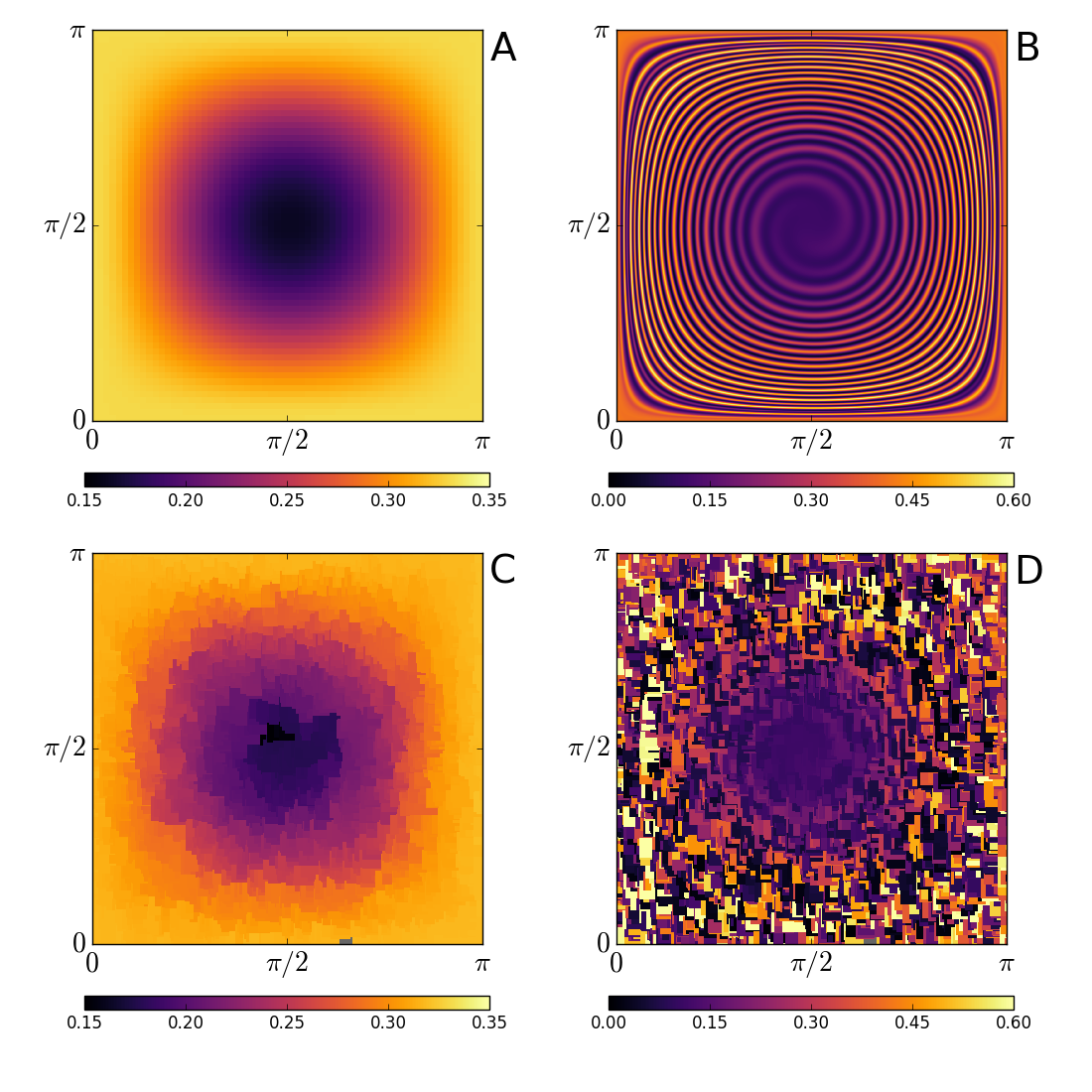} 
\par\end{centering}

\caption{\label{fig:Eulerian_vs_Lagrangian_fields}Field $c_{2}$ at time $t=100$.
One quarter of the whole domain is shown. A) pseudo--spectral scheme
on a $128\times128$ points grid. B) pseudo--spectral scheme on a
$4096\times4096$ points grid. C) Lagrangian scheme with $128^{2}$
particles and a diffusion matching that of A). D) Lagrangian scheme
with $128^{2}$ particles and a diffusion matching that of D). All
cases use the initial conditions (\ref{eq:initial_conditions_B}).}
\end{figure}

In this particular example, because of the quadratic nonlinearity,
the same amount of resource $c_{1}$ is consumed faster if it is spatially
concentrated than if it is spread out on a larger surface but at lower
concentrations. Thus smaller diffusivities, which better preserve
the concentration peaks of the resource, yield a faster growth of
the spatially averaged field $c_{2}$. In other words, they yield
a higher productivity of the consumer. 

One might then be lead to hope that, just as unresolved turbulence
can be usefully approximated by effective diffusion terms, in the
same way effective reaction terms should be sought, representing the
large--scale effects of the small--scale chemistry, with parameters
tuned as a function of the resolution of the model. Here we give an
example showing that this hope is unlikely to be fulfilled: we just
change the initial conditions (\ref{eq:initial_conditions_A}) with
\begin{equation}
c_{1}(x,y,0)=\left(\sin\left(\frac{x}{2}\right)\sin\left(\frac{y}{2}\right)\right)^{4},\quad c_{2}(x,y,0)=\left(\cos\left(\frac{x}{2}\right)\cos\left(\frac{y}{2}\right)\right),^{4}\label{eq:initial_conditions_B}
\end{equation}
and repeat the same calculations described above. Because the resource
and the consumer are now initially segregated into two nearly non--overlapping
blobs, larger diffusivities bring in contact the resource and the
consumer more quickly. As a result, we obtain the opposite effect
as before: the growth of the spatially averaged consumer is fastest
at the lowest resolution, and declines as the resolution is increased
(Figure \ref{fig:ConsumerProductivity}B). Thus, hypothetical effective
reaction terms intended to reproduce at low resolution the results
obtained at highest resolution with the chemistry (\ref{eq:simple_chemistry})
should achieve the no small feat of adjusting the productivity that
they yield not just to the resolution, but to the initial conditions,
too.

The diffusively--coupled Lagrangian scheme, having a diffusivity tunable
independently of the resolution, is not affected by these problems,
and reproduces fairly well with just $128^{2}$ particles the results
of the pseudo--spectral code using the same strengths of the diffusive
coupler as those used for Figure \ref{fig:ConsumerProductivity}A. 

The four panels of Figure \ref{fig:Eulerian_vs_Lagrangian_fields}
show the field $c_{2}$ at time $t=100$ as computed by the pseudo--spectral
scheme with 128 and 4096 grid points (panels A, B), and by the Lagrangian
scheme (panels C, D) with diffusivities matching those of the pseudo--spectral
calculations. The Lagrangian solutions are visualized by plotting
partially overlapping colored squares centered at the particles' positions,
rather than by resampling the solution on a regular grid. This choice
makes evident that the Lagrangian solution in panel D), reproduces
the same range of fluctuations as the solution in the panel B), even
though it obviously cannot resolve the fine structures created by
the advective dynamics.

The rest of the paper is organized as follows: in the following section
we describe the diffusive couplers; in section \ref{sec:Results}
we compare the results obtained through our Lagrangian methods against
known exact solutions or numerical solutions obtained with a pseudo--spectral
code at much higher resolution; in section \ref{sec:Implementation-details}
we briefly discuss how to efficiently implement the methods; finally
some concluding remarks are offered in section \ref{sec:Discussion-and-conclusions}.

\section{Diffusive couplers\label{sec:Diffusive-couplers}}

We are not going to attempt a discretization of the Laplacian operator:
evaluating the second derivatives of a field on a set of randomly
distributed points and then devising a numerical scheme that satisfies
mass conservation and the maximum principle would be quite challenging.
Of the two methods that we propose, the first the discrete counterpart
of a convolution with the heat kernel; the second represents diffusive
processes as exchanges of mass among nearby particles. Both methods
have free parameters, which determine the strength of the diffusive
effects. More precisely, they determine the rate at which the variance
of a scalar field is dissipated. In section \ref{sub:Advection-and-diffusion}
we give an objective, quantitative way to attach an effective diffusivity
to a given set of parameters.

We feel that the first coupler has a more mathematically elegant formulation.
However, it requires an iterative procedure to converge, which may
make it slow. The second coupler is little more than a recipe to destroy
variance, but its computational cost scales linearly with the number
of particles.

The problem of diffusively coupling the Lagrangian particles in (\ref{eq:ODEs})
is greatly simplified if one takes a fractional step approach \cite[§17.2]{LeVeque02}.
The reaction and advection terms are solved by integrating the ODEs
(\ref{eq:ODEs}) from time $t$ to time $t+\tau$, then a separate
diffusive step, which solves the heat equation, is performed. Here,
for notational simplicity, we illustrate the methods for the case
of one scalar field. The generalization of the methods to the $n$
scalar fields of the full PDEs (\ref{eq:the_problem}) is straightforward.

\subsection{First coupler\label{sub:First-coupler}}

In place of a discretized form of the heat equation, we seek a discretized
form of its solution; the latter, for a scalar field $c$, is given
by the following convolution integral 
\begin{equation}
c(\boldsymbol{x},t+\tau)=\int_{\Omega}k(\boldsymbol{x},\boldsymbol{y},\tau)c(\boldsymbol{y},t)\,\mathrm{d}\boldsymbol{y}\label{eq:convolution}
\end{equation}
where the kernel $k$ is the fundamental solution of the heat equation
in the domain $\Omega$ subject to the desired boundary conditions.
In $\mathbb{R}^{d}$ the kernel is 
\begin{equation}
k(\boldsymbol{x},\boldsymbol{y},\tau)=\left(\frac{1}{4\pi D\tau}\right)^{\frac{d}{2}}\exp\left(-\frac{\left\Vert \boldsymbol{x}-\boldsymbol{y}\right\Vert ^{2}}{4D\tau}\right)\label{eq:heat_kernel}
\end{equation}
where $D$ is the diffusion coefficient of the heat equation.

Given $M$ points $\boldsymbol{x}_{1},\ldots,\boldsymbol{x}_{M}$
in $\Omega$, let $W_{ij;\tau}$ be the elements of a matrix representing
a discrete counterpart of the convolution (\ref{eq:convolution})
evaluated at the points $\boldsymbol{x}_{i}$, $\boldsymbol{x}_{j}$
and across a time interval $\tau$. By analogy with the properties
of the kernel (\ref{eq:heat_kernel}), we shall assume $W$ to be
a non--negative, symmetric matrix. The simplest discretization of
the convolution (\ref{eq:convolution}) is given by 
\begin{equation}
c_{i}(t+\tau)=\sum_{j=1}^{M}W_{ij;\tau}c_{j}\label{eq:discrete_convolution}
\end{equation}
where we use the shorthands $c_{i}=c(\boldsymbol{x}_{i},t)$ and $c_{i}(t+\tau)=c(\boldsymbol{x}_{i},t+\tau)$.
If 
\begin{equation}
\sum_{j=1}^{M}W_{ij;\tau}=1,\label{eq:row_normalization}
\end{equation}
that is, each column of $W$ sums to $1$, then the expression (\ref{eq:discrete_convolution})
is just a weighted average of all the concentration values $\{c_{i}\}$.
Therefore, it satisfies the maximum principle in the form:

\begin{equation}
\min_{i=1,\ldots,M}\{c_{i}\}\le c_{i}(t+\tau)\le\max_{i=1,\ldots,M}\{c_{i}\}.\label{eq:discrete_max_principle}
\end{equation}
If each row of $W$ sums to $1$, i.e. 
\begin{equation}
\sum_{i=1}^{M}W_{ij;\tau}=1\label{eq:column_normalization}
\end{equation}
then the expression (\ref{eq:discrete_convolution}) satisfies the
conservation of mass in the form

\begin{equation}
\sum_{i=1}^{M}c_{i}(t+\tau)=\sum_{j=1}^{M}\left(\sum_{i=1}^{M}W_{ij;\tau}\right)c_{j}=\sum_{j=1}^{M}c_{j}.\label{eq:discrete_mass_conservation}
\end{equation}
Thus, if the discrete kernel $W$ is a \emph{doubly--stochastic} matrix
\cite{Sinkhorn-Knopp67}, i.e. it satisfies both (\ref{eq:row_normalization})
and (\ref{eq:column_normalization}), then the discrete model (\ref{eq:discrete_convolution})
obeys both the maximum principle and the conservation of mass.

Let us now describe how to construct such a discrete kernel $W$.
Initially we define a crude discretization of the exact kernel (\ref{eq:heat_kernel})
as follows 
\begin{equation}
K_{ij;\tau}=\begin{cases}
\exp\left(-\frac{\left\Vert \boldsymbol{x}_{i}-\boldsymbol{x}_{j}\right\Vert ^{2}}{4\mathcal{D}\tau}\right), & \left\Vert \boldsymbol{x}_{i}-\boldsymbol{x}_{j}\right\Vert <m\sqrt{2\mathcal{D}\tau}\\
0, & \left\Vert \boldsymbol{x}_{i}-\boldsymbol{x}_{j}\right\Vert \ge m\sqrt{2\mathcal{D}\tau}
\end{cases}\label{eq:kernel_discretization}
\end{equation}
where the nominal diffusivity $\mathcal{D}$ must be intended as a
free parameter. The kernel $K$ has a cut--off determined by $m$,
also a free parameter, to avoid computing the negligible contribution
of pairs of particles too far away from each other. Because $K$ is
not, in general, a doubly--stochastic matrix, we need to find a doubly--stochastic
surrogate of $K$.

The problem of rescaling a given matrix into a doubly--stochastic
one is named \emph{balancing}, and dates back to the 1930s. Since
then, a large number of applications has been solved by resorting
to the balance of matrices (see, e.g., \cite{Knight-Ruiz12} for a
rich list of examples).

We say that a matrix $K$ can be balanced if there exist two diagonal
matrices, $\text{diag}(\boldsymbol{a})$ and $\text{diag}(\boldsymbol{b})$,
such that 
\begin{equation}
W=\text{diag}(\boldsymbol{a})K\,\text{diag}(\boldsymbol{b})\label{eq:balancing}
\end{equation}
is doubly--stochastic. The fundamental theorem addressing this problem
for non--negative matrices is due to Sinkhorn and Knopp \cite{Sinkhorn-Knopp67}.
Starting from any vector $\boldsymbol{a}_{0}$ with positive elements,
they propose the following iteration: 
\begin{equation}
\boldsymbol{b}_{k+1}=\left(K^{T}\boldsymbol{a}_{k}\right)^{-1};\quad\boldsymbol{a}_{k+1}=\left(K\boldsymbol{b}_{k}\right)^{-1}\label{eq:SK_algorithm}
\end{equation}
where the reciprocal is intended to be applied element--wise. Their
theorem then states that the process converges to a doubly--stochastic
matrix of the form (\ref{eq:balancing}) with $\boldsymbol{a}=\lim_{k\to\infty}\boldsymbol{a}_{k}$,
$\boldsymbol{b}=\lim_{k\to\infty}\boldsymbol{b}_{k}$, if $K$ has
\emph{total support}. A matrix $K$ is said to have total support
if every positive entry in $K$ can be permuted into a positive diagonal
with a column permutation. Under the conditions of the theorem the
balancing is unique: $K$ can be turned into one and only one doubly--stochastic
matrix by means of multiplication by diagonal matrices (which are
themselves unique up to a scalar factor). 

Our crude discretization of the Gaussian kernel, the matrix (\ref{eq:kernel_discretization}),
has total support, because it is symmetric and has a positive main
diagonal. Therefore, if $K_{ij}$ is a non--zero element, then the
column permutation that swaps column $i$ with column $j$ brings
to the main diagonal $K_{ij}$, $K_{ji}$, and no other element; the
main diagonal thus remains positive. We can then define the discrete
convolution kernel $W$ that appears in (\ref{eq:discrete_convolution})
as the balancing of $K$. For our purposes it is important to note
that $K$ and $W$ have the same pattern of zeros, therefore the particle
pairs coupled by $W$ are all and only those coupled by $K$.

\subsection{Second coupler\label{sub:Second-coupler}}

A way to represent small-scale irreversible mixing processes is suggested
by physical intuition, along the following heuristic argument, similar
to those used in \cite{Haertel09,Bracco09}. When two fluid particles
happen to be close enough, they will exchange some portion of their
mass, and, thus, of their advected scalars. Let $q_{ij}\ge0$ be the
mass fraction exchanged between the $i-$th and the $j-$th particle,
which are assumed to have the same mass. This fraction may be a function
of the distance $\left\Vert \boldsymbol{x}_{i}-\boldsymbol{x}_{j}\right\Vert $
and may be assumed to be zero when the distance exceeds some fixed
threshold. Thus the concentration of the scalar $c$ after a diffusion
step at the position of the $i-$th particle will be 
\begin{equation}
c_{i}(t+\tau)=c_{i}-\sum_{j=1}^{M}q_{ij}c_{i}+\sum_{j=1}^{M}q_{ij}c_{j}\label{eq:Exchange mixing}
\end{equation}
where the first sum represents the losses to other particles, and
the second sum represents the gains from other particles. The above
expression can be re-arranged as 
\begin{equation}
c_{i}(t+\tau)=\left(1-\sum_{j=1}^{M}q_{ij}\right)c_{i}+\left(\sum_{j=1}^{M}q_{ij}\right)\overline{c}_{i}\label{eq:Exchange mixing maximum principle}
\end{equation}
where the overline denotes the weighted average $\overline{c}_{i}=\sum_{j=1}^{M}q_{ij}c_{j}/\sum_{j=1}^{M}q_{ij}$.
If 
\begin{equation}
0\le\sum_{j=1}^{M}q_{ij}\le1\label{eq:Echange mixing maximum principle condition}
\end{equation}
equation (\ref{eq:Exchange mixing maximum principle}) shows that
$c_{i}(t+\tau)$ is a linear interpolation between $c_{i}$ and $\overline{c}_{i}$,
and therefore the maximum principle is satisfied.

In addition, it is straightforward to verify that $\sum_{i}c_{i}(t+\tau)=\begin{subarray}{c}
\sum_{i}\end{subarray}c_{i}$, and therefore the expression (\ref{eq:Exchange mixing}) conserves
mass.

As exchange fraction we shall use 
\begin{equation}
q_{ij}=\begin{cases}
\frac{p}{\left(4\pi\mathcal{D}\tau\right)^{\frac{d}{2}}}\,\exp\left(-\frac{\left\Vert \boldsymbol{x}_{i}-\boldsymbol{x}_{j}\right\Vert ^{2}}{4\mathcal{D}\tau}\right), & \left\Vert \boldsymbol{x}_{i}-\boldsymbol{x}_{j}\right\Vert <m\sqrt{2\mathcal{D}\tau}\\
0, & \left\Vert \boldsymbol{x}_{i}-\boldsymbol{x}_{j}\right\Vert \ge m\sqrt{2\mathcal{D}\tau}
\end{cases}\label{eq:exchange_fraction}
\end{equation}
where $p$, $\mathcal{D}$ and $m$ are free parameters and $d$ is
the dimensionality of the space. This particular choice is loosely
suggested by the fact that if the scalar field carried by the $i-$th
particle at time $t$ were represented by a delta function, a diffusion
process having diffusivity $\mathcal{D}$, after a time $\tau$ would
spread out the scalar over the whole domain with a resulting concentration
proportional to $\exp\left(-\left\Vert \boldsymbol{x}_{i}-\boldsymbol{x}_{j}\right\Vert ^{2}/(4\mathcal{D}\tau)\right)$.
The cut--off for large distances is also physically motivated: the
small-scale, unresolved advective motions that this diffusion process
is supposed to represent, cannot occur at an arbitrarily large speed;
therefore, in a finite time $\tau$ only particles closer than some
threshold length may exchange mass.

Special care must be taken in choosing $p$ small enough as to enforce
the condition (\ref{eq:Echange mixing maximum principle condition}).
A useful rule of thumb is: 
\begin{equation}
\frac{p}{(4\pi\mathcal{D}\tau)^{\nicefrac{d}{2}}}<\frac{1}{N(m\sqrt{2\mathcal{D}\tau})},\label{eq:rule_of_thumb}
\end{equation}
where $N(h)$ is the average number of particles that fall into a
sphere of radius $h$.

\subsection{Boundary conditions}

So far we have discussed the diffusive couplers as if the computational
domain were unbounded. When the domain is limited, any condition enforced
along its boundaries is reflected in the kernel $k$ appearing in
the convolution solution (\ref{eq:convolution}), which ceases to
be a simple Gaussian function.

In the case of periodic boundary conditions, the kernel is an infinite
sum of Gaussians, one for each of the periodic images. For example,
on the segment $[0,2\pi)$ the kernel is
\begin{equation}
k(x,y,\tau)=\sum_{n\in\mathbb{Z}}\frac{1}{\sqrt{4\pi D\tau}}\exp\left(-\frac{\left(x-y+2n\pi\right)^{2}}{4D\tau}\right).\label{eq:1D_periodic_heat_kernel}
\end{equation}
If $m\sqrt{2D\tau}<\pi$, and we accept to approximate to zero the
exponential when its argument is larger than or equal to $m$ (as
we do in (\ref{eq:kernel_discretization}) and in (\ref{eq:exchange_fraction})),
then only one term gives a non--zero contribution in the sum. This
shows that the expressions (\ref{eq:kernel_discretization}) and (\ref{eq:exchange_fraction})
remain valid for periodic boundary conditions, provided that the norms
$\left\Vert \boldsymbol{x}_{i}-\boldsymbol{x}_{j}\right\Vert $ which
appear in those expressions are considered as the minimum distance
in the periodic domain between the particle $i$ and the particle
$j$.

Another common boundary condition prescribes that the flux of tracers
across any portion of the boundary has to be zero. When no particle
is seeded outside of the domain, this condition is automatically enforced
by both the diffusive couplers presented here. There is, however,
a pitfall that needs to be brought to light. This is most easily illustrated
in a one--dimensional domain. Let us consider the half--line $[0,\infty)$.
If we impose no--flux (a.k.a Neumann) boundary conditions at $x=0$,
then the heat kernel is 
\begin{equation}
k(x,y,\tau)=\frac{1}{\sqrt{4\pi D\tau}}\left[\exp\left(-\frac{\left(x-y\right)^{2}}{4D\tau}\right)+\exp\left(-\frac{\left(x+y\right)^{2}}{4D\tau}\right)\right].\label{eq:no-flux_heat_kernel}
\end{equation}
This can be deduced by imposing an even symmetry to the initial condition
which extends the problem to the whole line, and then restricting
the solution back to the half--line. The even symmetry enforces the
boundary condition. This implies that the points at $x>0$ do exchange
fluxes across the boundary with their mirror images at $x<0$, but
do so as to keep equal to zero the net flux at $x=0$. If these virtual
fluxes across the boundary are not taken into account, then, in proximity
of the boundaries, the diffusivity of the scalar field is underestimated,
even though the no--flux boundary condition is still correctly enforced.
A solution to this problem might consist in using ghost particles
strategically placed outside the domain so as to represent an even--symmetric
field across it. In more than one dimension, this would be relatively
straightforward only for straight boundaries, and would quickly escalate
to a challenging problem for boundaries of arbitrary shape. However,
the contribution of the mirror images is important only within a distance
of $O(\sqrt{2D\tau})$ from the boundary. In high--Péclet number,
under--resolved simulations, this distance would be comparable to
or smaller than the inter--particle distance. We thus feel that attempting
to fix this issue may not be worth the effort. In the following when
we mention ``no--flux boundary condition'' we refer to the straightforward
case in which no ghost particles are used.

In the test cases we have not used the Dirichlet boundary condition.
However we anticipate no difficulties in implementing this condition
by distributing particles along the boundary and fixing their concentrations
to a prescribed value. The same considerations about mirror images
and ghost particles, subject to the appropriate symmetry, apply to
this case as well.

\section{Results\label{sec:Results}}

\subsection{Advection and diffusion\label{sub:Advection-and-diffusion}}

\begin{figure}
\begin{centering}
\includegraphics[width=0.7\columnwidth]{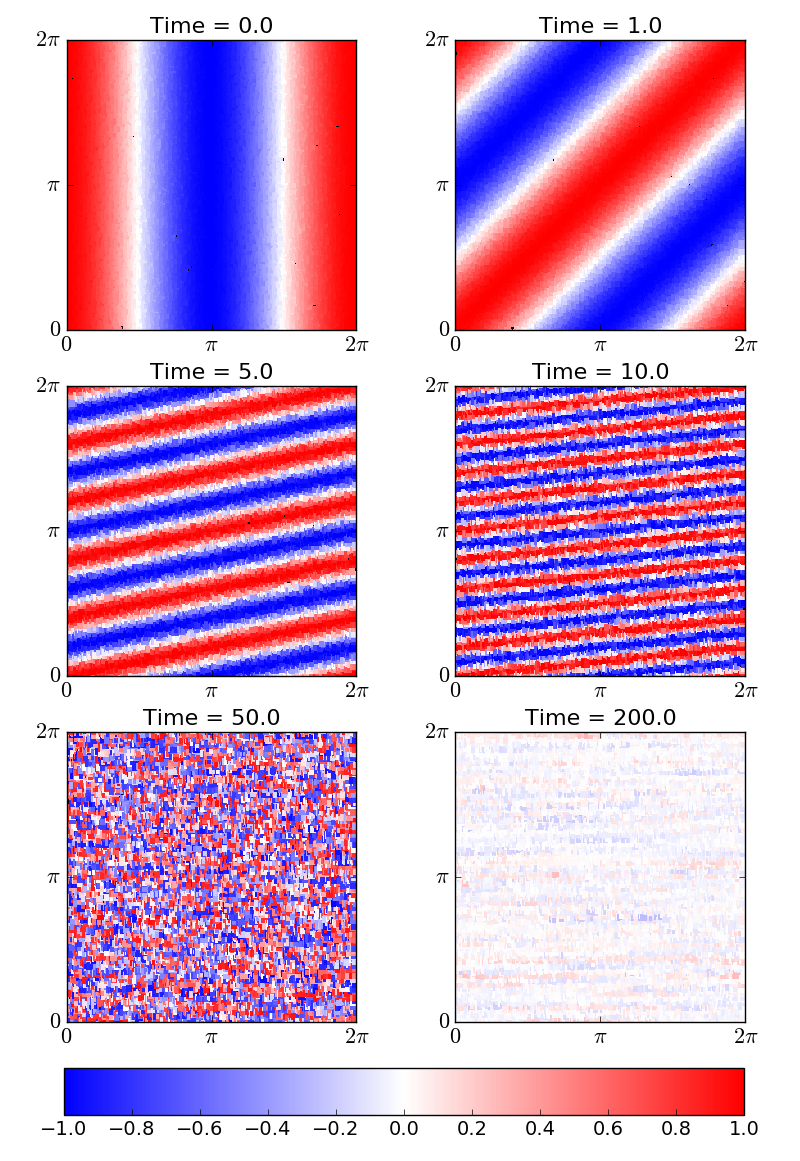}
\par\end{centering}

\caption{\label{fig:RY-time-evolution}Numerical solution of (\ref{eq:rhines-young},\ref{eq:rhines-young_IC})
using the first coupler (§\ref{sub:First-coupler}). The parameters
of the discretized kernel (\ref{eq:kernel_discretization}) are $d=2$,
$m=8$, $\sqrt{2\mathcal{D}\tau}=\pi/512$, $\tau=0.1$. The second
coupler, with the parameters of Figure \ref{fig:RY-dissipation_rate},
produces visually indistinguishable results.}
\end{figure}

A first test for the diffusive couplers introduced in the previous
section is to compare their performance for advection--diffusion problems
in cases in which small--scale structures are progressively formed
and eventually become under--resolved. An analytically--solvable,
well--known, but non trivial test case is the following \cite{Rhines-Young83}:
\begin{equation}
\frac{\partial c}{\partial t}+y\frac{\partial c}{\partial x}=D\nabla^{2}c\label{eq:rhines-young}
\end{equation}
with initial condition 
\begin{equation}
c(x,y,0)=\cos(x).\label{eq:rhines-young_IC}
\end{equation}
In a domain vertically unbounded and horizontally periodic with period
of $2\pi$, the problem (\ref{eq:rhines-young},\ref{eq:rhines-young_IC})
has the exact solution 
\begin{equation}
c(x,y,t)=e^{-D\left(t+\frac{t^{3}}{3}\right)}\cos\left(x-yt\right)\label{eq:rhines-young_solution}
\end{equation}
which develops arbitrarily high wavenumbers in the $y-$direction
as times progresses due to the tipping over of the tracer streaks
operated by the shearing flow (Figure \ref{fig:RY-time-evolution}).
 Multiplying (\ref{eq:rhines-young}) by $c$, averaging, and using
(\ref{eq:rhines-young_solution}) after an integration by parts, one
finds the following explicit expression for the rate of dissipation
of scalar variance 
\begin{equation}
-\frac{d}{dt}\left\langle \frac{c^{2}}{2}\right\rangle =D\left\langle \left|\nabla c\right|^{2}\right\rangle =\frac{D}{2}\left(1+t^{2}\right)e^{-2D\left(t+\frac{t^{3}}{3}\right)}.\label{eq:rhines-young_dissipation_rate}
\end{equation}
Where the angular brackets denote a spatial average over one horizontal
period and an arbitrary vertical length. 

\begin{figure}
\begin{centering}
\includegraphics[width=0.6\columnwidth]{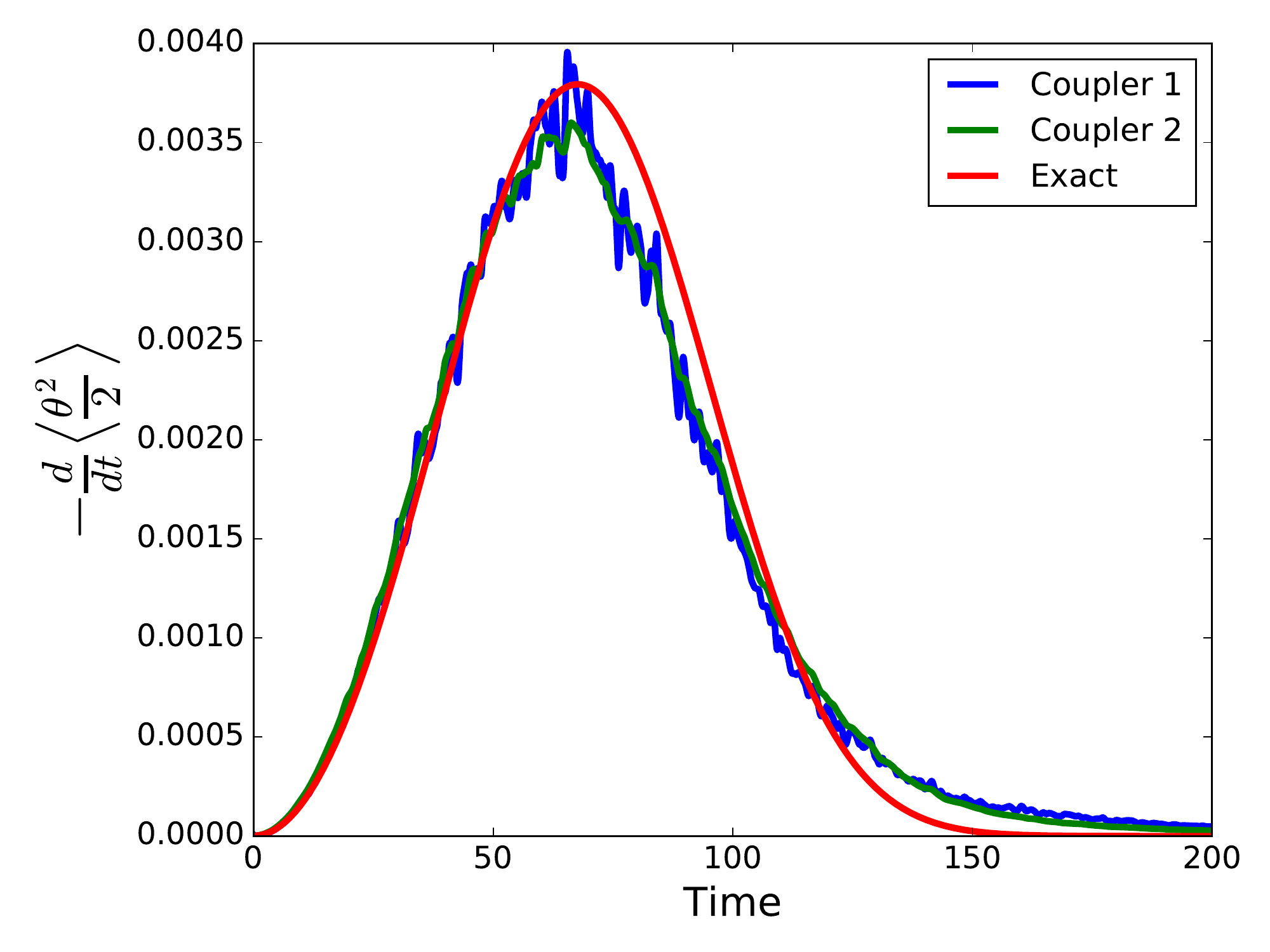}
\par\end{centering}

\caption{\label{fig:RY-dissipation_rate}Rate of dissipation of scalar variance
for the problem (\ref{eq:rhines-young},\ref{eq:rhines-young_IC}).
Blue curve: results from the numerical simulation of Figure \ref{fig:RY-time-evolution}.
Green curve: results using the second coupler (§\ref{sub:Second-coupler}),
with parameters $p=1.38\cdot10^{-5}$, $m=4$, $\sqrt{2\mathcal{D}\tau}=\pi/256$,
$\tau=0.1$ in the exchange fraction (\ref{eq:exchange_fraction}).
Red curve: expression (\ref{eq:rhines-young_dissipation_rate}) with
$D=3.23\cdot10^{-6}$.}
\end{figure}

In Figure \ref{fig:RY-dissipation_rate} this expression is compared
with the results obtained using the two couplers discussed in sec.
(\ref{sec:Diffusive-couplers}).
 The numerical computations use the domain $[0,2\pi)\times[-\pi,3\pi]$,
periodic in $x$ and with no--flux boundary conditions in $y$. The
number of particles is $128\times256$. The averages are computed
in the central part of the domain, shown in Figure \ref{fig:RY-time-evolution}.
The left--hand side of (\ref{eq:rhines-young_dissipation_rate}) is
then computed from the particles' concentrations. The value of the
diffusivity $D$ in the right--hand side of (\ref{eq:rhines-young_dissipation_rate})
is least--squares fitted to the numerical results. The fit extends
from the beginning of the simulation up to the time of maximum dissipation.
The value of the parameter $p$ in the second coupler is tuned in
order to match the fitted value of $D=3.23\ldots\cdot10^{-6}$ obtained
with the first coupler with at least two significant digits.

The match with the exact dissipation rate becomes inaccurate at later
times, because when the stripes become under--resolved the tracer
variance is aliased to lower wave numbers, and thus it is not damped
as quickly as it should have been. The advantages of the Lagrangian
approach, however, should become clear by observing that with a pseudo--spectral
code at a comparable resolution, the lowest diffusivity must be $D\approx10^{-3}$
in order to avoid significant spurious oscillations. With that diffusivity
the dissipation rate peaks at time $t\approx10$ instead than $t\approx70$,
by which time the streaks have all but disappeared. With the Lagrangian
approach we are able to obtain a qualitatively correct shape of the
dissipation curve corresponding to a diffusivity that is three orders
of magnitude smaller.

In fact, for each choice of the parameters, we can define the \emph{effective
diffusivity} of the method as the value $D$ in the right-hand side
of (\ref{eq:rhines-young_dissipation_rate}) that best fits the growing
part of the numerical dissipation curve. This value, in general, does
not coincide with the \emph{nominal diffusivity} $\mathcal{D},$ which
appears in (\ref{eq:kernel_discretization}) and (\ref{eq:exchange_fraction})
and depends on the parameters as we shall discuss below.

\begin{figure}
\begin{centering}
\includegraphics[width=0.6\columnwidth]{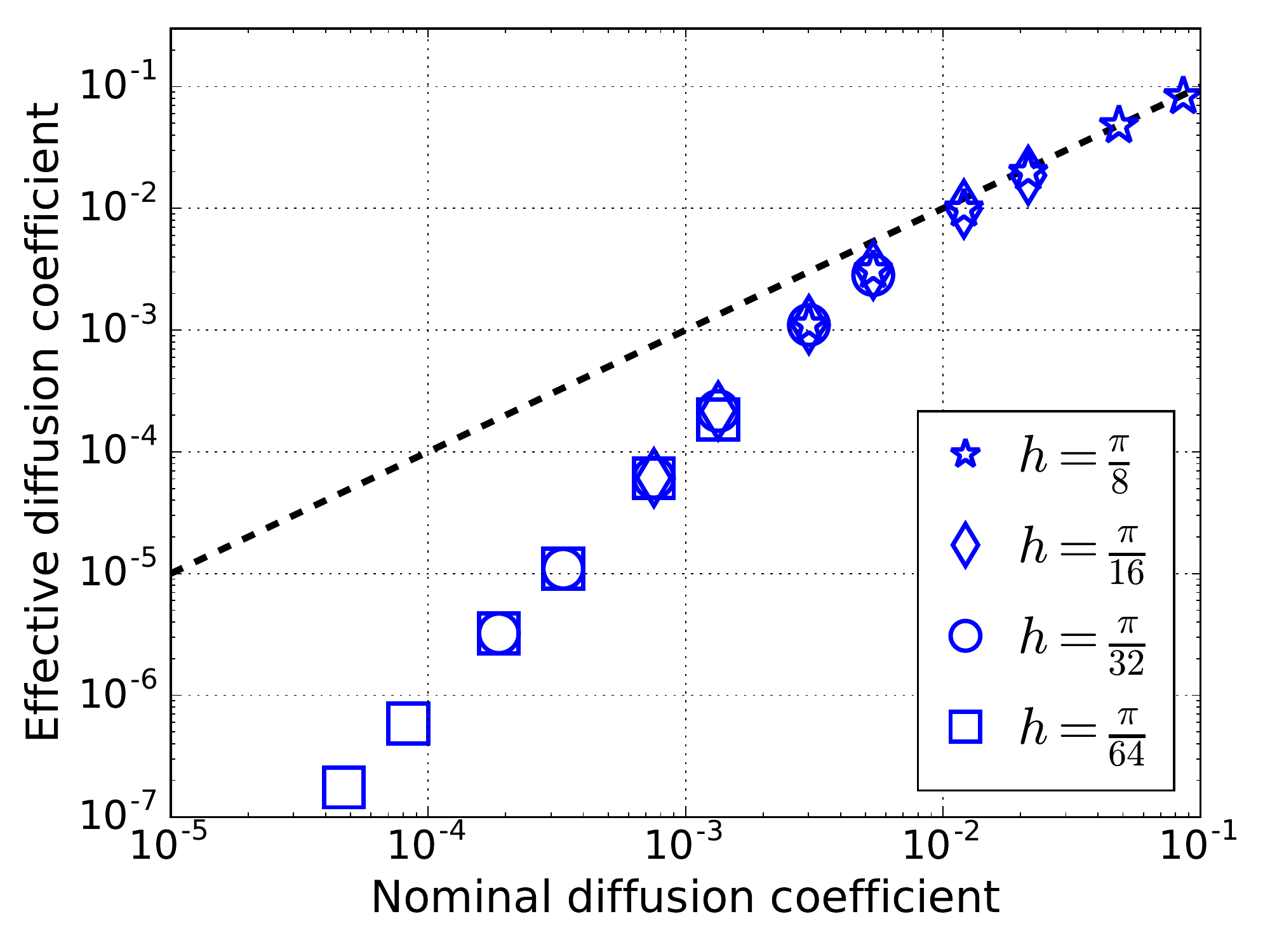}
\par\end{centering}

\caption{\label{fig:SK_nominal_vs_effective}Effective diffusivity $D$ as
a function of nominal diffusivity $\mathcal{D}$ for the first coupler
(§\ref{sub:First-coupler}). Different symbols correspond to different
values of the cut--off radius $h$. Different points with the same
symbol correspond to different values of $m$. The nominal diffusivity
is then given by (\ref{eq:nominal_diffusivity}). The black dashed
line is the identity $D=\mathcal{D}$.}
\end{figure}

Using the first coupler, in the discrete kernel (\ref{eq:kernel_discretization})
we set the cut--off radius $m\sqrt{2\mathcal{D}\tau}=h$ to be $h=\nicefrac{\pi}{8},\,\nicefrac{\pi}{16},\,\nicefrac{\pi}{32},\,\nicefrac{\pi}{64}$.
For each of these values we consider $m=3,4,6,8,12,16$. Fixing the
value of the time step (we use $\tau=0.1$) the nominal diffusivity
is then determined as
\begin{equation}
\mathcal{D}=\frac{h^{2}}{2\tau m^{2}}.\label{eq:nominal_diffusivity}
\end{equation}
Figure \ref{fig:SK_nominal_vs_effective} shows the effective diffusivity
as a function of the nominal diffusivity for the above values of $h$
and $m$.
 Points that have the same $\nicefrac{h}{m}$ ratio yield nearly the
same effective diffusivity. In other words, for fixed $\mathcal{D}$,
the effective diffusivity is fairly insensitive to the cut-off radius
$h$, even when this is so small that only very few particles are
involved: when $h=\nicefrac{\pi}{64}$ only $\pi$ particles, on average,
fall within a disc of radius $h$.

At high nominal diffusivities, the effective diffusivity nearly coincides
with the nominal one: $D(\mathcal{D})\approx\mathcal{D}$. At low
nominal diffusivities the effective diffusivity appears to be proportional
to the square of the nominal one: $D(\mathcal{D})\propto\mathcal{D}^{2}$.
Further tests suggest that the constant of proportionality scales
as the square root of the particle density, and that the switch between
the two regimes occurs when the standard deviation $\sqrt{2\mathcal{D}\tau}$
of the discrete kernel (\ref{eq:kernel_discretization}) is of the
same order of magnitude as the average distance between nearest particles.
We did not further investigate the reasons of this change of slope
and postpone an in-depth examination of the issue to a further work.

\begin{figure}
\begin{centering}
\includegraphics[width=0.6\columnwidth]{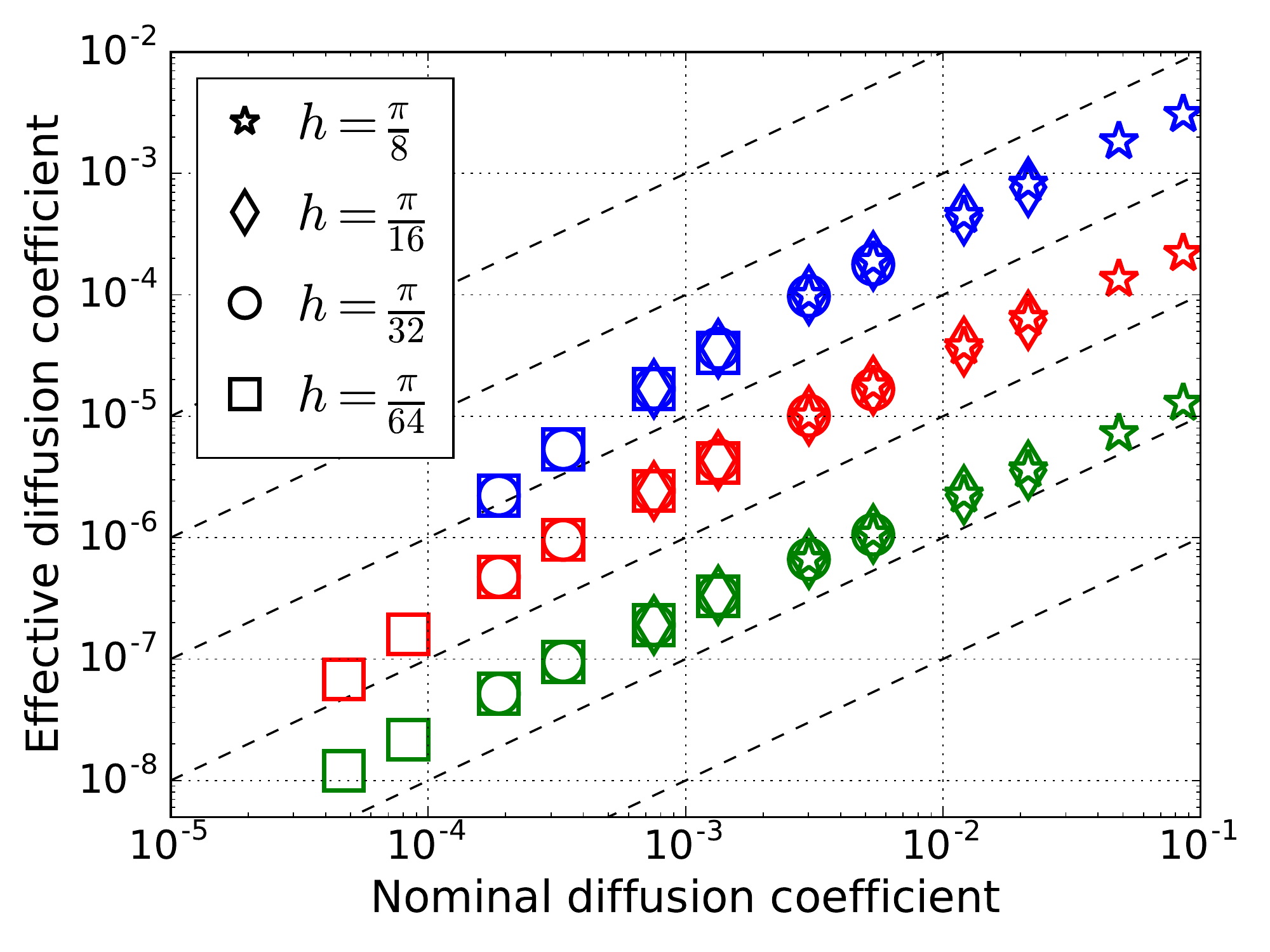}
\par\end{centering}

\caption{\label{fig:noSK_nominal_vs_effective}Effective diffusivity $D$ as
a function of nominal diffusivity $\mathcal{D}$ for the second coupler
(§\ref{sub:Second-coupler}). Symbols have the same meaning as in
Figure \ref{fig:SK_nominal_vs_effective}. Blue markers refer to computations
with $p=10^{-4}$, red to $p=10^{-5}$, green to $p=10^{-6}$. The
black dashed lines are the functions $D=10^{n}\mathcal{D}$, with
$n=-5,-4,\ldots,0$.}
\end{figure}

Figure \ref{fig:noSK_nominal_vs_effective} shows 
 the effective diffusivity obtained with the second coupler as a function
of the coupler's parameters appearing in the exchange fraction (\ref{eq:exchange_fraction}).
The markers relative to $h=\nicefrac{\pi}{64}$, $m=12,16$ are absent,
because with those parameters the condition (\ref{eq:Echange mixing maximum principle condition})
does not hold: the method violates mass conservation and blows up.

The cut-off radius is determined as specified above for the first
coupler, and the expression (\ref{eq:nominal_diffusivity}) for the
nominal diffusivity still holds. As in that case, the effective diffusivity
is fairly insensitive to the cut-off radius $h$ when the ratio $\nicefrac{h}{m}$
is kept fixed. In contrast with the first coupler, the effective diffusivity
appears to be roughly proportional to the nominal one across the whole
range of diffusivities that we have tested. The effective diffusivity
also appears to be roughly proportional to the parameter $p$. 

The effective diffusivity of the second coupler also depends on the
density of the particles. If, keeping all other parameters the same,
we double the average number of particles that fall within a disk
of radius $h$, we find, from (\ref{eq:Exchange mixing}) and (\ref{eq:exchange_fraction}),
that the average mass exchanged on a time step by each particle with
its neighbors doubles. Thus the effective diffusivity is proportional
to the particle density.

\subsection{Reaction and diffusion}

The methods described in the present work are designed for cases in
which the Péclet numbers are extremely high. However, it cannot be
excluded that some geophysical flows may, occasionally, be characterized
by less extreme Péclet numbers. It is thus of interest to verify what
may be the performance of the methods when the advection terms are
not dominant over the diffusion ones. In the limit of zero Péclet
numbers, the equations (\ref{eq:the_problem}) reduce to reaction--diffusion
equations. Even though we are not proposing our methods for this class
of problems, we found informative to use one of them as a test case. 

Here we will consider the well-known Fisher--Kolmogorov--Petrovskii--Piskunov
equation, namely
\begin{equation}
\frac{\partial c}{\partial t}=D\nabla^{2}c+c(1-c).\label{eq:KPP}
\end{equation}
For non-negative $c$, this equation describes the propagation of
fronts joining a stable ($c=1$) and an unstable ($c=0$) region (e.g.
\cite{Murray07}§13.2). There exist solutions with fronts propagating
at any speed $V\ge2\sqrt{D}$. However, for a very large class of
initial conditions, in particular those whose derivative has compact
support, the propagation speed is the minimal one\cite{vanSaarlos03}:
$V=2\sqrt{D}$.

When the function $c$ assumes negative values the solution generally
blows--up to minus infinity in a finite time. It is thus important
to avoid numerical solution methods that generate spurious oscillations.
In particular, this may be a problem when the diffusion coefficient
is small, because the thickness of the front is also proportional
to $\sqrt{D}$. Thus, low diffusivities imply high gradients in the
traveling front.

We produce a numerical approximation of (\ref{eq:KPP}) by uniformly
random seeding $128^{2}$ particles in the square $[0,2\pi]\times[0,2\pi]$.
We use no--flux boundary condition. Initially, all particles have
a concentration of zero, except those having a coordinate $x<0.2$,
whose concentration is set to one. We then advance the solution with
time steps of length $\tau=0.1$ by alternating one of the diffusive
couplers of sec. \ref{sec:Diffusive-couplers} and the evaluation
of the exact solution of the equation $\dot{c}=c(1-c)$.

\begin{figure}
\begin{centering}
\includegraphics[width=0.6\columnwidth]{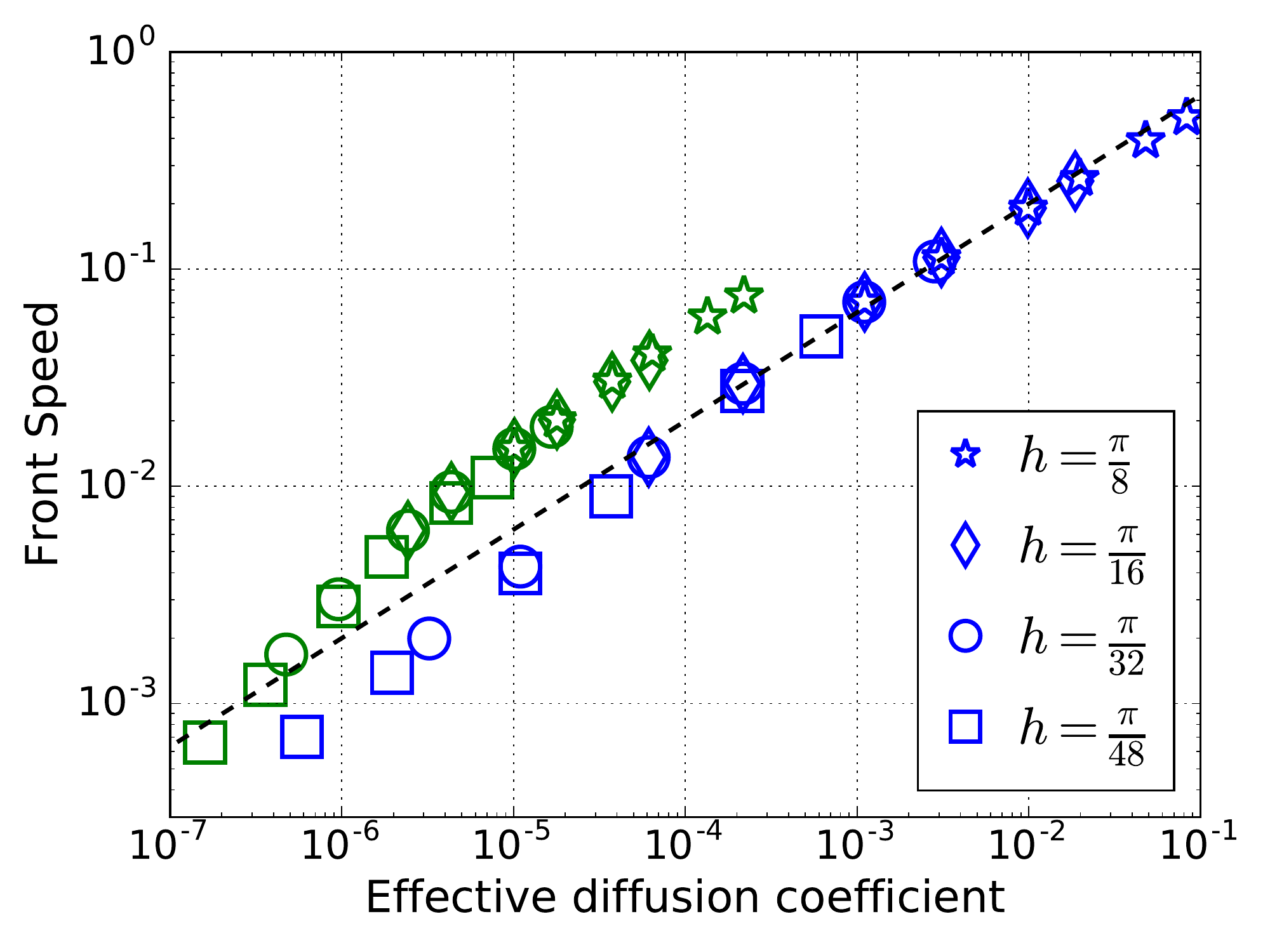}
\par\end{centering}

\caption{\label{fig:V_vs_effective_diffusivity}Speed of propagation of the
front in the solution of equation (\ref{eq:KPP}) as a function of
the effective diffusivity. Symbols have the same meaning as in Figure
\ref{fig:SK_nominal_vs_effective}. Blue markers refer to computations
using the first coupler (sec.\ref{sub:First-coupler}) and the the
green ones to computations using the second coupler (\ref{sub:Second-coupler})
with $p=10^{-5}$. The black dashed line is the theoretical speed
$V=2\sqrt{D}$.}
\end{figure}
In Figure \ref{fig:V_vs_effective_diffusivity} we plot the propagation
speed of the front as a function of the effective diffusivity of the
method, evaluated as detailed in the previous subsection. The first
coupler gives the best results, while the second coupler overestimates
the speeds by about a factor 2.5. With both couplers the front propagation
speed appears to be proportional to the square root of the diffusivity,
as in the exact solution, except at very low diffusivities, where
the front speed declines somewhat faster than the exact scaling. This
excessive slow--down is in qualitative agreement with what was found
in a stochastically forced version of equation (\ref{eq:KPP}). The
primary effect of the random forcing was that of damping the leading
tail of the propagating front, thus slowing it down \cite{Doering03}.
We speculate that the random arrangement of the particles may play
the role of the stochastic forcing.

The front is well--resolved only at the lowest diffusivities. When
$D\approx10^{-3}$ the thickness of the front becomes comparable with
the interparticle distance. Thus most of the results of Figure \ref{fig:V_vs_effective_diffusivity}
refer to runs in which the front is poorly resolved or not resolved
at all. When the front is not resolved, the separation between the
region where $c=1$ and $c=0$ appears as a jagged line, with meanders
of characteristic size determined by the interparticle distance.

We could not run this test case with a cut--off radius $h=\pi/64$,
because this length results to be smaller than the percolation threshold:
due to the random inhomogeneities in the distribution of the particles,
after a short transient, no particle with concentration zero is found
at a distance less than $h$ from a particle with concentration higher
than zero, thus the front stops propagating. In figure \ref{fig:V_vs_effective_diffusivity},
we used $h=\pi/48$, instead. This elucidates the disadvantage of
not having a velocity field stirring the particles: although Poissonian
random gaps in the distribution of particles exist even in the presence
of a stirring velocity field, they open and close as time progresses,
rather than remaining static, and are thus far less damaging, as the
results of the other tests should clearly illustrate.

\subsection{Advection, reaction and diffusion at different Damköhler numbers}

We now return to the simple resource--consumer model (\ref{eq:simple_chemistry})
to test the performance of the Lagrangian couplers when the Damköhler
number is changed. Here we do so by letting the reaction rate assume
the values $r=0.04,\,0.2,\,1,\,5$, while keeping in all cases the
same velocity field, which is defined by the following streamfunction
\begin{equation}
\psi(x,y,t)=\left[(n\bmod2)\sin(x+\phi_{n})-\left(1-(n\bmod2)\right)\sin(y+\phi_{n})\right]\label{eq:renovating_flow}
\end{equation}
where $n=\left\lfloor t\right\rfloor $ (the largest integer smaller
than $t$), ``$\bmod$'' denotes the remainder of the integer division,
and $\phi_{n}$ is a uniformly random phase chosen in $[0,2\pi)$.
This is an example of a ``random renewing flow'' (see e.g. \cite{Childress_Stretch-twist-fold}§11.1)
which is very effective at mixing an advected scalar field. The characteristic
spatial scale of the flow is constant, but an advected field is subject
to a continuous process of stretching and folding that produces a
cascade of progressively smaller scales.

Our benchmarks are numerical solutions of the problem (\ref{eq:the_problem})
with the chemistry (\ref{eq:simple_chemistry}) and the velocity field
induced by (\ref{eq:renovating_flow}), solved on a uniform grid with
$4096^{2}$ nodes, on the doubly--periodic domain $[0,2\pi)\times[0,2\pi)$,
with a Fourier--Galerkin pseudo--spectral code, and a diffusion coefficient
$D=\nicefrac{0.003}{32^{2}}\approx2.9\cdot10^{-6}$. A slightly larger
diffusivity was used than in the computations of Figure \ref{fig:ConsumerProductivity}
at the same resolution, because at higher reaction rates the solution
develops higher gradients in the concentration fields. We thus have
tuned $D$ so as to obtain a solution free of spurious oscillations
at $r=5$, and we have kept that value for all the reaction rates.
We use both the uniform consumer initial condition (\ref{eq:initial_conditions_A})
and the non overlapping blobs initial condition (\ref{eq:initial_conditions_B}).

\begin{figure}
\begin{centering}
\includegraphics[width=1\columnwidth]{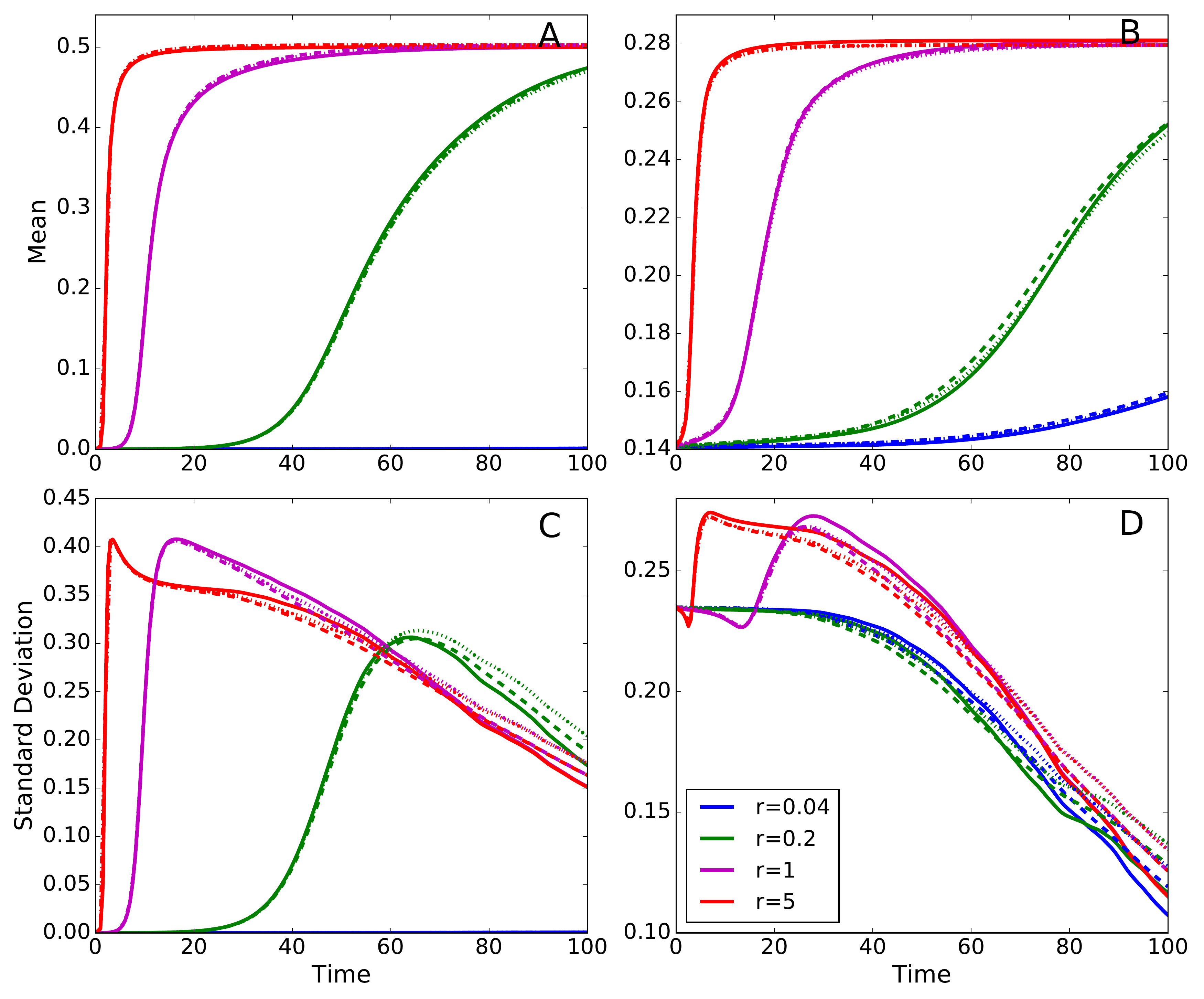}
\par\end{centering}

\caption{\label{fig:renowating_wave_varying_r}Mean (panels A,B) and standard
deviation (panels C,D) of the consumer concentration field as a function
of time using the chemistry (\ref{eq:simple_chemistry}) and the stirring
field (\ref{eq:renovating_flow}). Panels A,C refer to the initial
conditions (\ref{eq:initial_conditions_A}); panels B,D to the initial
conditions (\ref{eq:initial_conditions_B}). Different colors denote
different reaction rates, as specified in the legend of panel D. Solid
lines refer to results obtained with a pseudo--spectral code on a
grid with $4096^{2}$ nodes. Dotted and dashed lines refer to the
Lagrangian method with $128^{2}$ particles and respectively, the
coupler of section \ref{sub:First-coupler} and of section \ref{sub:Second-coupler}. }
\end{figure}
Against the benchmark we compare the results obtained using the Lagrangian
method with the couplers of section \ref{sec:Diffusive-couplers}.
For the first coupler we use a cut--off radius $h=\pi/64$ and $m=5.8$.
For the second coupler we use $h=\pi/32$, $m=4$, $p=10^{-5}$. In
both cases $128^{2}$ particles were used, the time step is $\tau=0.1$
and the ODEs (\ref{eq:ODEs}) are solved with the standard fourth--order
Runge--Kutta scheme. The results are summarized in Figure \ref{fig:renowating_wave_varying_r}.

Both Lagrangian methods reproduce very well the time evolution of
the mean of the chemical fields, and reasonably well their standard
deviation. If measured with the criterion of section \ref{sub:Advection-and-diffusion},
the parameters above yield an effective diffusivity slightly higher
( $D\approx1.1\cdot10^{-5}$) than the diffusivity of the pseudo--spectral
code (the criterion suggests $m\approx8$ for the first coupler and
$m\approx7.5$ for the second). If the effective diffusivity matches
that of the pseudo--spectral code, in the later stages of the simulation
the standard deviation remains too high and decays at a rate clearly
slower than in the pseudo--spectral benchmark. This occurs because,
as stirring cascades the chemical tracers to unresolved small scales,
the variance relative to those scales is aliased back to larger scales,
where it is damped at an incorrect, lower rate. Using an ad-hoc higher
effective diffusivity initially gives a slight underestimation of
the standard deviation and, later on, a slight overestimation, while
producing what we consider to be an acceptable approximation of a
dynamics that requires a resolution 32 times higher to be fully resolved.

\section{Implementation details\label{sec:Implementation-details}}

An efficient implementation of the diffusive couplers of section \ref{sec:Diffusive-couplers}
requires a fast algorithm for finding all the particles falling within
a distance $h$ from any given particle. This \emph{fixed--radius
near neighbors search} is a classic problem in computational geometry.
For arbitrary distribution of points, it can be solved by arranging
the points in tree data structures such as quad--trees or Kd--trees
(see e.g. \cite{ComputationalGeometry} chap. 5, 14). The use of trees
leads to algorithms with a computational cost of $O\left(M\log(M)\right)$,
where $M$ is the number of points. When, as in our case, the particles
are uniformly distributed, it is more convenient to use a lattice
and hashing method, which has a computational cost of $O(M)$ \cite{Bentley77}. 

The computational domain is overlaid with a regular lattice with square
meshes of size $h$. To each mesh is assigned a unique index. For
simplicity we use row--order indexing, although the Z--order indexing
might improve cache efficiency. The particles are kept in a list,
sorted according to the index of the mesh that contains each particle,
which is easily computed from the particle's position. The sorting
is performed by means of the \emph{counting} algorithm (e.g. \cite{Introduction_to_algorithms}§8.2),
which does not use pairwise comparisons, and has a complexity of $O(M)$.
A hash table associates each mesh index with the first particle in
the sorted list having that index. Thus, accessing all particles within
the same mesh is an $O(1)$ operation, because each mesh, on average,
contains the same number of particles, due to their uniform distribution.
To find all the particles within a distance $h$ from a given one,
one needs to compute the distance of the given particle from all the
particles in the same mesh and in some of the adjacent meshes (three
in 2D or four in 3D). After each time step, the particle list is sorted
again, and the hash table is updated. If the size of the mesh $h$
is decreased as the number of particles $M$ increases in such a way
as to maintain constant the average number of particles in each mesh,
then the fixed--radius neighbor search problem is solved in $O(M)$
time. We did not attempt yet to produce a parallel version of our
prototype code. However we don't expect to face unusual difficulties
or harsh performance penalties by pursuing a straightforward domain
partitioning strategy, in which each processor takes care of a contiguous
block of meshes.

In the case of the coupler of section \ref{sub:Second-coupler}, the
computation of the exchange fraction (\ref{eq:exchange_fraction})
only increases the prefactor in the asymptotic scaling of the fixed--radius
neighbor search. The overall algorithm is thus $O(M)$. 

In the case of the coupler of section \ref{sub:First-coupler}, an
analysis of the computational cost is more complicated, because it
needs to take into account the cost of balancing the discrete kernel
(\ref{eq:kernel_discretization}). The analysis of balancing algorithms
is still an open problem, and we settled for the venerable Sinkhorn--Knopp
algorithm only because it is extremely simple to implement. An assessment
of the performances and of the relative merits of balancing algorithms,
in particular on distributed--memory parallel architectures, is beyond
the scope of this paper, and might become the subject of a future
work.

\section{Discussion and conclusions\label{sec:Discussion-and-conclusions}}

In this paper we have investigated the viability of Lagrangian numerical
methods to approximate the solution of advection--reaction--diffusion
equations in cases where it is impossible to resolve all the scales
of motion, as is commonplace for biogeochemical problems.

The methods consist in alternating a purely Lagrangian step that solves
the advection--reaction part of the equations with the method of characteristics,
with a diffusive step that couples the particles moving along the
characteristic lines of the advection--reaction problem. Two such
couplers have been proposed. One amounts to a discrete version of
the convolution with a Gaussian kernel, the other prescribes the exchanges
between nearby particles of small portions of the mass carried by
each. In both cases the resulting scheme conserves mass, respects
the maximum principle and allows to tune the diffusivity down to zero,
where the couplers have no effect, and the method of characteristics
is recovered.

We have carried several tests comparing the methods against exact
solutions of advection--diffusion and reaction--diffusion problems,
and against fully resolved numerical solutions of advection--reaction--diffusion
problems obtained using a pseudo--spectral method run at significantly
higher resolution than that of the Lagrangian code. In all cases the
results have been fairly good, except in the case of the reaction--diffusion
test, where the lack of an advection term that stirs the particles
hampers the performance of the method. However, even in this unfavorable
case, the methods are able to recover in a roughly correct way the
main features of the solution and their scaling as a function of the
diffusivity.

Of course, when it is impossible to resolve all the spatial scales
present in the solution, no method should be considered as completely
satisfactory, and it is very likely that special cases could be found
where it would perform far from well. For example, we don't expect
our Lagrangian method to perform brilliantly in reproducing the propagation
of chemical fronts stirred by steady cellular flows. The speed of
those fronts critically depends on an accurate description of the
tails of the tracer distribution in proximity of the hyperbolic stagnation
points at the cell boundaries \cite{Tzella15}. When the spatial structures
are severely under resolved those tails are not reproducible, and
the resulting speed is then unlikely to be correct. On the other hand,
chemical fronts of that kind are not present in the oceans, and stagnation
points, albeit present, are not steady; typical ocean mixing processes
involve shearing, or stretching and folding dynamics, and in those
cases our approach seems to be satisfactory. 

This paper does not suggest that our Lagrangian methods are competitive
with, or even comparable to, a fully resolved numerical solution obtained
with an Eulerian method, but rather that, by allowing to control the
diffusivity independently of the resolution, the Lagrangian methods
offer, when resolution can't be further increased, a much better compromise
than equally unresolved Eulerian methods. In this respect, diffusive
couplers like those presented here could be seen more as a subgrid--scale
parameterization of sorts, rather than as a discretization of a diffusion
operator such as the Laplacian that appears in (\ref{eq:the_problem}).

While we believe that the present work is a successful proof--of--concept,
some additional steps will be required in order to incorporate it
into a realistic ocean circulation model. The current prototype implementation
needs to be extended to three spatial dimensions and to distributed--memory
parallel architectures. In the present form the couplers only represent
homogeneous and isotropic diffusive processes. In ocean models, anisotropy
is necessary, at least in the vertical direction, and the possibility
to allow for spatially--dependent diffusivities is desirable. Finally,
the existing Eulerian parameterizations for the sources and sinks
of tracers, due to interactions with the bottom, with the air, and
through river run--off must be adapted to the Lagrangian framework.
These goals will probably be easier to achieve by modifying the coupler
of section \ref{sub:Second-coupler} where subgrid--scale fluxes are
represented explicitly and locally as exchanges of mass between particles.
They may be more demanding with the coupler of section \ref{sub:First-coupler},
which requires the balancing of a matrix, a process that involves
all particles simultaneously, even when the discretized kernel that
couples them has a cut--off at a finite distance.

\subsection*{Acknowledgements}

Part of this work was funded by Università del Salento through ``Progetto
5xmille per la Ricerca''. We have benefited from discussions with
Zouhair Lachkar, Olivier Paulis and Marcello Vichi. We are indebted
to Clare Eayrs, Marina Levi and Shafer Smith who read and commented
earlier drafts of the paper.

\end{document}